\documentclass[showpacs,preprintnumbers,amsmath,amssymb,twocolumn]{revtex4}
\usepackage{graphicx}
\usepackage{dcolumn}
\usepackage{bm}
\begin{document}

\title{\bf\large{Pion Condensation in Baryonic Matter: from Sarma Phase to Larkin-Ovchinnikov-Fudde-Ferrell Phase}}
\author{Lianyi He\footnote{Email address: hely04@mails.tsinghua.edu.cn}, Meng Jin\footnote{Email address: jin-m@mail.tsinghua.edu.cn
} and Pengfei Zhuang\footnote{Email address:
zhuangpf@mail.tsinghua.edu.cn}}
\affiliation{Physics Department,
Tsinghua University, Beijing 100084, China}

\begin{abstract}
We investigated two pion condensed phases in the frame of the two
flavor Nambu--Jona-Lasinio model at finite baryon density: the
homogeneous and isotropic Sarma phase and inhomogeneous and
anisotropic Larkin-Ovchinnikov-Fudde-Ferrell(LOFF) phase. At small
isospin chemical potential $\mu_I$, the Sarma state is free from
the Sarma instability and magnetic instability due to the strong
coupling and large enough effective quark mass. At large $\mu_I$,
while the Sarma instability can be cured via fixing baryon density
$n_B$ to be nonzero, its magnetic instability implies that the
LOFF state is more favored than the Sarma state. In the
intermediate $\mu_I$ region, the stable ground state is the Sarma
state at higher $n_B$ and LOFF state at lower $n_B$.
\end{abstract}

\date{\today}
 \maketitle

\section {Introduction}
\label{s1}
Recently, the study on  Quantum Chromodynamics (QCD) phase
structure is extended to finite isospin density\cite{son}. The
physical motivation to study QCD at finite isospin density and the
corresponding pion superfluidity is related to the investigation
of compact stars, isospin asymmetric nuclear matter and heavy ion
collisions at intermediate energies. In early studies on dense
nuclear matter and compact stars, it has been suggested that
charged pions are condensed at sufficiently high isospin
density\cite{migdal,sawyer,camp,kaplan}.

While the perturbation theory of QCD can well describe the
properties of new QCD phases at extremely high temperature and
density, the study on the phase structure at moderate temperature
and density depends on lattice QCD calculation and effective
models with QCD symmetries. The lattice simulation at finite
isospin chemical potential\cite{kogut} shows that there is a phase
transition from normal phase to pion superfluidity phase at a
critical isospin chemical potential which is about the pion mass
in the vacuum, $\mu_I^c \simeq m_\pi$. The QCD phase structure at
finite isospin density is also investigated in low energy
effective models, such as the Nambu--Jona-Lasinio (NJL) model
applied to quarks\cite{toub,bard,frank,he-1,he-2,warringa} which
is simple but enables us to see directly how the dynamic mechanism
of isospin symmetry breaking operates. In the frame of this model,
it is analytically proved\cite{he-1,he-2} that the critical
isospin chemical potential for pion superfluidity is exactly the
pion mass in the vacuum, $\mu_I^c = m_\pi$, and this relation is
independent of the model parameters and regularization scheme. In
addition, near the phase transition point, the chiral and pion
condensates calculated in this model are in good agreement with
the lattice simulation\cite{kogut}. However, up to now, most of
the studies in lattice and model calculations at quark level is
done at zero baryon chemical potential, and the conditions of
charge neutrality and beta equilibrium which are required in
physical systems such as neutron stars can not be achieved. It is
therefore necessary to study the effect of finite baryon chemical
potential on the pion condensation in the frame of NJL model at
quark level. Recently, this issue has been discussed in chiral
limit\cite{ebert}.

In a pion superfluid at zero baryon chemical potential, the quark
and antiquark of a condensed pair have the same isospin chemical
potential and in turn the same Fermi surface (Strictly speaking,
the term "Fermi surface" is only meaningful in weakly coupled
fermi gas). When a nonzero baryon chemical potential is turned on,
it can be regarded as a Fermi surface mismatch between the quark
and antiquark. The Cooper pairing between different fermions with
mismatched Fermi surfaces was discussed many years ago. In early
investigation of superconductivity in an external magnetic field,
Sarma found a spatially uniform state\cite{sarma} where there
exist gapless fermion excitations. However, compared with the
fully gapped BCS state, the gapless state is energetically
unfavored, which is called Sarma instability\cite{sarma}.
Recently, such a spatially uniform ground state prompted new
interest theoretically due to the studies on fermion pairing with
large mass difference\cite{liu}, two-component fermionic atom gas
with density difference\cite{pao,son2}, isospin asymmetric nuclear
matter with neutron-proton pairing\cite{np}, and neutral color
superconducting quark matter\cite{CSC1,CSC2,CSC3,CSC4}. It is now
generally accepted that the Sarma instability can be cured in some
cases, such as the system with long-range interaction where charge
neutrality is required\cite{CSC2,CSC3} and the system with proper
finite range interaction between the two species of fermions with
large mass difference\cite{forbes}. In the Sarma state, the
dispersion relation of one branch of the quasi-particles has two
zero points at momenta $p_1$ and $p_2$ which behave as two
effective Fermi surfaces, and at these two points it needs no
energy for quasi-particle excitations. Since the system contains
both a superfluid Fermi liquid component in the region
$p<p_1,p>p_2$ and a normal Fermi liquid component in the region
$p_1<p<p_2$, the Sarma state is now also called breached pairing
state or interior gap state.

Unfortunately, it is found that the Sarma state suffers negative
superfluid density\cite{wu} or negative Meissner mass
squared\cite{CI1,he-3} which indicates a magnetic instability, and
the LOFF state\cite{LOFF1,LOFF2} with spatially non-uniform
condensate is more energetically favored than the Sarma
state\cite{ACI1,ACI3,he-4}. In the study of isospin asymmetric
color superconducting quark matter, isospin asymmetric nuclear
matter and atomic fermion gas with density difference, the LOFF
phase is proposed to be the ground state and widely
investigated\cite{LOFF3,LOFF4,LOFF5,LOFF6,LOFF7}. However, it is
recently argued that, the Sarma phase will be the stable ground
state of asymmetric fermion superfluid in the Bose-Einstein
condensation(BEC) region\cite{pao,son2,smit}. In relativistic
fermion superfluids, the same conclusion is obtained when the
coupling is strong enough and the fermions are heavy
enough\cite{kita}. While the Sarma state may not be realized in
color superconductivity, since the effective quark mass is very
small at large baryon density, it is probably stable in a pion
condensate matter with nonzero baryon density, since the effective
quark mass is large enough near the phase transition point of pion
condensation, i.e., around the isospin chemical potential
$\mu_I=m_\pi$. In this paper, we study the Sarma state and LOFF
state in pion condensed matter at finite baryon density in the
frame of NJL model.

\begin{figure}[!htb]
\begin{center}
\includegraphics[width=6cm]{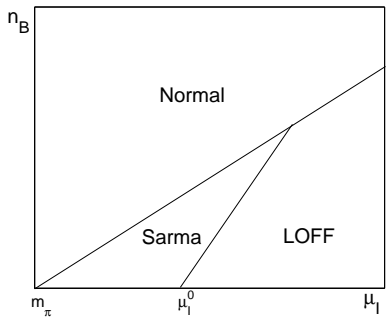}
\caption{The schematic phase diagram of pion condensation in the
$\mu_I-n_B$ plane. \label{fig1}}
\end{center}
\end{figure}
The qualitative phase diagram of pion condensation at finite
baryon density we obtained in this paper is illustrated in
Fig.\ref{fig1} in the isospin chemical potential and baryon
density plane. The pion condensation can exist when the baryon
density is not very high, otherwise the system will be in normal
phase without pion condensation because of the too strong
mismatch. The critical baryon density for the transition from pion
condensation to normal phase increases with the isospin chemical
potential. The pion condensation starts at the critical value
$\mu_I^c=m_\pi$ and is separated into two phases at another
critical value $\mu_I^0$. At small $\mu_I$, the Sarma phase is
free from the Sarma instability and magnetic instability due to
the strong coupling and large quark mass, it is therefore the
stable ground state. At large $\mu_I$, the Sarma phase suffers the
Sarma instability and magnetic instability. While the Sarma
instability can be cured via fixed nonzero baryon density, the
magnetic instability implies that the LOFF phase is more favored
than the Sarma phase at large $\mu_I$. The schematic phase diagram
of pion condensation at nonzero baryon density we obtained here is
similar to the phase diagram of cold polarized fermi gas proposed
by Son and Stephanov\cite{son2}, and the isospin chemical
potential here plays the same role of the coupling in atomic fermi
gas there.

The paper is organized as follows. In Section \ref{s2}, we review
the formalism of pion condensation in the NJL model at finite
isospin chemical potential and baryon chemical potential. In
section \ref{s3}, we investigate the property of pion condensation
at zero baryon chemical potential and show that there is a BEC-BCS
crossover when the isospin chemical potential increases. In
section \ref{s4}, we evaluate the Sarma pion condensed phase and
discuss the gapless quasiparticle spectrum. In section \ref{s5},
we study the thermodynamic stability of the Sarma phase. In
section \ref{s6}, we calculate the response of the Sarma phase to
an external electromagnetic current and isospin current and
discuss the dynamical stability of the Sarma phase. In section
\ref{s7}, we construct the LOFF pion condensed phase and discuss
when it is more favored than the Sarma phase. We summarize in
section \ref{s8}.

\section {NJL Model at Finite $\mu_I$ and $\mu_B$}
\label{s2}
In this section, we review the formalism of NJL model at finite
temperature and isospin and baryon chemical potentials\cite{he-2}.
Since the isospin chemical potential which triggers pion
condensation is large, $\mu_I\geq m_\pi$, we neglect the
possibility of diquark condensation which is favored at large
baryon chemical potential and small isospin chemical potential.
The key quantity to describe the thermodynamics of a system at
finite baryon and isospin densities is the partition function
defined by
\begin{equation}
Z(T,\mu_I,\mu_B,V)= \text {Tr}\ e^{-\beta(H-\mu_B B-\mu_I I_3)},
\end{equation}
where $V$ is the volume of the system, $\beta$ the inverse
temperature $\beta = 1/T$, $\mu_B$ and $\mu_I$ are the baryon and
isospin chemical potentials, and $B$ and $I_3$ the conserved
baryon number and isospin number operators. If we take only quark
field $\psi$ as elementary field of the system, the operators can
be expressed as
\begin{equation}
B = \frac{1}{3}\int d^{3}{\bf x}\bar{\psi}\gamma_{0}\psi\ , \ \ \
\ I_{3} = \frac{1}{2}\int d^3{\bf
x}\bar{\psi}\gamma_{0}\tau^{3}\psi,
\end{equation}
the factors $1/3$ and $1/2$ reflect the fact that $3$ quarks make
a baryon and quark's isospin quantum number is $1/2$. In the
imaginary time formalism of finite temperature field theory, the
partition function can be represented as
\begin{equation}
Z(T,\mu_I,\mu_B,V)=\int[d\bar{\psi}][d\psi][dA]e^{-\int_{0}^{\beta}d\tau\int
d^{3}{\bf x}\left({\cal
L}+\bar{\psi}\hat{\mu}\gamma_{0}\psi\right)},
\end{equation}
where ${\cal L}$ is the Lagrangian density describing the system,
and $\hat{\mu}$ the quark chemical potential matrix in flavor
space $\hat{\mu}= diag(\mu_u,\mu_d)$ with the $u$ and $d$ quark
chemical potentials
\begin{equation}
\mu_u = \frac{\mu_B}{3}+\frac{\mu_I}{2}\ , \ \ \ \ \mu_d =
\frac{\mu_B}{3}-\frac{\mu_I}{2}.
\end{equation}
For convenience, we also use the notation $\mu\equiv\mu_B/3$ in
the following.

The flavor $SU(2)$ NJL Lagrangian density is defined as
\begin{equation}
{\cal L} =
\bar{\psi}\left(i\gamma^{\mu}\partial_{\mu}-m_0\right)\psi
+G\left[\left(\bar{\psi}\psi\right)^2+\left(\bar{\psi}i\gamma_5{\bf
\tau}\psi\right)^2 \right]
\end{equation}
with scalar and pseudoscalar interactions corresponding to
$\sigma$ and ${\bf \pi}$ excitations. The Lagrangian density has
the symmetry $U_B(1)\bigotimes SU_I(2)\bigotimes SU_A(2)$
corresponding to baryon number symmetry, isospin symmetry and
chiral symmetry, respectively. However, at nonzero isospin
chemical potential, the isospin symmetry $SU_I(2)$ breaks down to
$U_I(1)$ global symmetry with the generator $I_3$ which is related
to the condensation of charged pions. At zero baryon chemical
potential, the Fermi surfaces of $u (d)$ and anti-$d (u)$ quarks
coincide and hence the condensate of $u$ and anti-$d$ quarks is
favored at sufficiently high $\mu_I
>0$ and the condensate of $d$ and anti-$u$ quarks is favored at
sufficiently high $\mu_I <0$. We introduce the chiral condensate,
\begin{equation}
\label{mf7}
\langle\bar{\psi}\psi\rangle = \sigma,
\end{equation}
and the pion condensates,
\begin{eqnarray}
\langle\bar{\psi}i\gamma_5\tau_+\psi\rangle &=&
\sqrt{2}\langle\bar{u}i\gamma_5d\rangle=\pi^+={\pi\over \sqrt
2}e^{i\theta},\nonumber\\
\langle\bar{\psi}i\gamma_5\tau_-\psi\rangle &=&
\sqrt{2}\langle\bar{d}i\gamma_5u\rangle=\pi^-={\pi\over \sqrt
2}e^{-i\theta}
\end{eqnarray}
with $\tau_{\pm} = \left(\tau_1 \pm i\tau_2\right)/\sqrt{2}$. A
nonzero condensate $\sigma$ means spontaneous chiral symmetry
breaking, and a nonzero condensate $\pi$ means spontaneous isospin
symmetry breaking. The phase factor $\theta$ related to the
condensates $\pi_+$ and $\pi_-$ indicates the direction of the
$U_I(1)$ symmetry breaking. If the ground state is uniform,
$\theta$ is a constant and it does not alter the physical result.
For convenience we choose $\theta=0$ in the following.

In mean field approximation the partition function is simplified
as
\begin{equation}
Z(T,\mu_I,\mu_B,V)=\int[d\bar{\psi}][d\psi]e^{-\int_{0}^{\beta}d\tau\int
d^{3}{\bf x}{\cal L}_{mf}}
\end{equation}
with the mean field Lagrangian density
\begin{equation}
{\cal L}
=\bar{\psi}\left[i\gamma^\mu\partial_\mu-m+\hat{\mu}\gamma_0
-i\Delta\tau_1\gamma_5\right]-G(\sigma^2+\pi^2)\psi,
\end{equation}
where $m=m_0-2G\sigma$ is the effective quark mass and
$\Delta=-2G\pi$ is the effective energy gap. From the mean field
Lagrangian density, we can derive the inverse quark propagator
matrix in flavor space as a function of quark momentum,
\begin{equation}
{\cal S}^{-1}(p)=\left(\begin{array}{cc} \gamma^\mu p_\mu+\mu_u\gamma_0-m & -i\gamma_5\Delta\\
-i\gamma_5\Delta & \gamma^\mu
p_\mu+\mu_d\gamma_0-m\end{array}\right).
\end{equation}
Using the formula listed in Appendix A, the thermodynamic
potential in mean field approximation,
\begin{equation}
\Omega=-{T\over V}\ln Z=G(\sigma^2+\pi^2)-{T\over V}\ln\det{{\cal
S}^{-1}}
\end{equation}
can be evaluated as a summation of four quasiparticle
contributions,
\begin{equation}
\Omega= G(\sigma^2+\pi^2)-6\sum_{i=1}^4\int {d^3 {\bf p}\over
(2\pi)^3}g(\omega_i(p)),
\end{equation}
where $\omega_i$ are the dispersions of the quasiparticles,
\begin{eqnarray}
&&\omega_1(p)=E_\Delta^-+\mu,\ \ \ \omega_2(p)=E_\Delta^--\mu,\nonumber\\
&&\omega_3(p)=E_\Delta^++\mu,\ \ \ \omega_4(p)=E_\Delta^+-\mu
\end{eqnarray}
with the definitions
\begin{eqnarray}
&&E_\Delta^\pm = \sqrt{(E_p^\pm)^2+\Delta^2}\ ,\nonumber\\
&&E_p^\pm=E_p\pm\mu_I/2,\ \ \ E_p = \sqrt{p^2+m^2},
\end{eqnarray}
and the function $g(x)$ is defined as $g(x)=x/2+T\ln(1+e^{-x/T})$.
The gap equations to determine the effective quark mass $m$ and
pion condensate $\Delta$ can be obtained by the minimum of the
thermodynamic potential,
\begin{equation}
\label{gap}
\frac{\partial\Omega}{\partial m}=0,\ \ \
\frac{\partial\Omega}{\partial\Delta}=0,\ \ \
{\partial^2\Omega\over \partial m^2}>0,\ \ \
{\partial^2\Omega\over\partial \Delta^2}>0.
\end{equation}
From the first order derivatives, we have
\begin{eqnarray}
\label{gap0} m-m_0 &=& 12G\int{d^3{\bf p}\over (2\pi)^3}{m\over
E_p} \Big[{E_p^-\over
E_\Delta^-}\left(1-f(\omega_1)-f(\omega_2)\right)\nonumber\\
&&+{E_p^+\over
E_\Delta^+}\left(1-f(\omega_3)-f(\omega_4)\right)\Big] ,\nonumber\\
\Delta &=& 12 G\Delta\int{d^3{\bf p}\over (2\pi)^3}\Big[{1\over
E_\Delta^-}\left(1-f(\omega_1)-f(\omega_2)\right)\nonumber\\
&&+{1\over E_\Delta^+}\left(1-f(\omega_3)-f(\omega_4)\right)\Big]
\end{eqnarray}
with the Fermi-Dirac distribution function $f(x) = 1/\left(
e^{x/T}+1\right)$. This group of gap equations is invariant under
the transformation $\mu_I\rightarrow-\mu_I$, and we can only
concentrate on the case $\mu_I>0$. Once $\Omega$ is known, the
thermodynamic functions such as the pressure $p$, the entropy
density $s$, the charge number densities $n_B$ and $n_I$, the
flavor number densities $n_u$ and $n_d$, the energy density
$\epsilon$, and the specific heat $c$ can be obtained by
thermodynamic relations.

To investigate the quark propagation in the pion condensed matter,
we should derive the explicit form of the quark propagator at mean
field level\cite{he-2},
\begin{equation}
{\cal S}(p)= \left(\begin{array}{cc} {\cal S}_{uu}(p)&{\cal S}_{ud}(p)\\
{\cal S}_{du}(p)&{\cal S}_{dd}(p)\end{array}\right)
\end{equation}
with the elements
\begin{eqnarray}
{\cal S}_{uu} &=&
{\left(p_0+\mu+E_p^-\right)\Lambda_+\gamma_0\over
(p_0-\omega_2)(p_0+\omega_1)}+ {\left(p_0+\mu-E_p^+\right)\Lambda_-\gamma_0\over (p_0-\omega_4)(p_0+\omega_3)},\nonumber\\
{\cal S}_{dd} &=&
{\left(p_0+\mu-E_p^-\right)\Lambda_-\gamma_0\over
(p_0-\omega_2)(p_0+\omega_1)}+ {\left(p_0+\mu+E_p^+\right)\Lambda_+\gamma_0\over (p_0-\omega_4)(p_0+\omega_3)},\nonumber\\
{\cal S}_{ud} &=& {-i\Delta \Lambda_+\gamma_5\over
(p_0-\omega_2)(p_0+\omega_1)}+ {-i\Delta \Lambda_-\gamma_5\over (p_0-\omega_4)(p_0+\omega_3)},\nonumber\\
{\cal S}_{du} &=& {-i\Delta \Lambda_-\gamma_5\over
(p_0-\omega_2)(p_0+\omega_1)}+ {-i\Delta \Lambda_+\gamma_5\over
(p_0-\omega_4)(p_0+\omega_3)},
\end{eqnarray}
where $\Lambda_\pm$ are the energy projectors
\begin{equation}
\Lambda_{\pm}({\bf p}) = {1\over 2}\left(1\pm{\gamma_0\left({\bf
\gamma}\cdot{\bf p}+m\right)\over E_p}\right).
\end{equation}

The occupation number of each kind of quarks, namely u, d, anti-u
and anti-d quarks, is useful for our discussion in the following.
They can be calculated from the positive and negative energy
components of the diagonal propagators ${\cal S}_{uu}$ and ${\cal
S}_{dd}$\cite{he-2},
\begin{eqnarray}
n_u^+(p) &=&
u_-^2f\left(\omega_2\right)+v_-^2f\left(-\omega_1\right),\nonumber\\
n_d^-(p) &=&
u_-^2f\left(\omega_1\right)+v_-^2f\left(-\omega_2\right), \nonumber\\
n_d^+(p) &=&
u_+^2f\left(\omega_4\right)+v_+^2f\left(-\omega_3\right), \nonumber\\
n_u^-(p) &=&
u_+^2f\left(\omega_3\right)+v_+^2f\left(-\omega_4\right),
\end{eqnarray}
where the symbols $+$ and $-$ in the notations $n_{u,d}^\pm$ stand
for quarks and antiquarks, respectively, and the coherent
coefficients $u_\pm^2$ and $v_\pm^2$ are defined as
\begin{eqnarray}
u_\pm^2
&=&\frac{1}{2}\left(1+\frac{E_p^\pm}{E_\Delta^\pm}\right),\ \
v_\pm^2 =\frac{1}{2}\left(1-\frac{E_p^\pm}{E_\Delta^\pm}\right).
\end{eqnarray}
The number density of each kind of quarks and the total baryon
number density are
\begin{eqnarray}
n_{u,d}^\pm &=& 6\int{d^3{\bf p}\over
(2\pi)^3}n_{u,d}^\pm(p),\\
n_B &=& 2\int{d^3{\bf p}\over
(2\pi)^3}\left[f\left(\omega_2\right)-f\left(\omega_1\right)+f\left(\omega_4\right)-f\left(\omega_3\right)\right].\nonumber
\end{eqnarray}

\section {BEC-BCS Crossover at Zero $\mu_B$}
\label{s3}
It has been argued both in effective theory and lattice simulation
that at finite but not very large isospin density and zero baryon
density, the QCD matter is a pure meson matter, i.e., a
Bose-Einstein condensate of charged pions. At ultra high isospin
density, the matter turns to be a Fermi liquid with
quark-antiquark cooper pairing\cite{son}. Therefore, there should
be a BEC to BCS crossover when the isospin chemical potential
increases. In this section we discuss the gap equations
(\ref{gap0}) at $T=\mu_B=0$ which has been investigated in
\cite{he-2} and point out that there are some signals of BEC to
BCS crossover in the NJL model at finite isospin density.

\begin{figure}[!htb]
\begin{center}
\includegraphics[width=6.5cm]{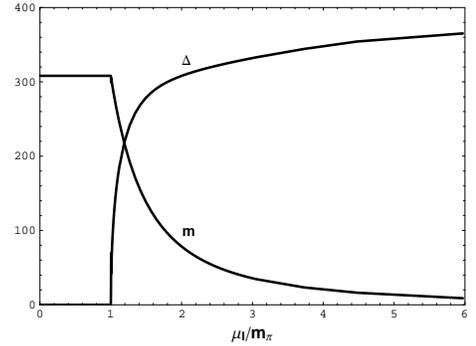}
\caption{The effective quark mass $m$ and the pion condensate
$\Delta$ as functions of $\mu_I/m_\pi$. \label{fig2}}
\end{center}
\end{figure}

Since the NJL model is non-renormalizable, we should employ a hard
three momentum cutoff $\Lambda$ to regularize the gap equations.
In the following numerical calculations, we take the current quark
mass $m_0=5$ MeV, the coupling constant $G=4.93$ GeV$^{-2}$ and
the cutof $\Lambda=653$ MeV. This group of parameters ensures the
pion mass $m_\pi=138$ MeV and the pion decay constant $f_\pi=93$
MeV in the vacuum. Numerical solution of the gap equations is
shown in Fig.\ref{fig2}. For $\mu_I<m_\pi$ the ground state is the
same as the vacuum and the isospin density keeps zero, while for
$\mu_I>m_\pi$ the pion condensate and isospin density become
nonzero. The critical isospin chemical potential $\mu_I^c$ is
exactly the vacuum pion mass $m_\pi$, and the phase transition is
of second order. Since the chiral symmetry is spontaneously broken
at small $\mu_I$ and almost restored at large $\mu_I$, the
effective quark mass $m$ plays an important role at small $\mu_I$
but can be neglected at large $\mu_I$. In fact, the effective
quark mass as a function of $\mu_I$ at $\mu_I>m_\pi$ can be well
described by\cite{he-2}
\begin{equation}
\frac{m(\mu_I)}{m(0)}\simeq\frac{\sigma(\mu_I)}{\sigma(0)}=\left(\frac{m_\pi}{\mu_I}\right)^2\
.
\end{equation}

\begin{figure}[!htb]
\begin{center}
\includegraphics[width=6.5cm]{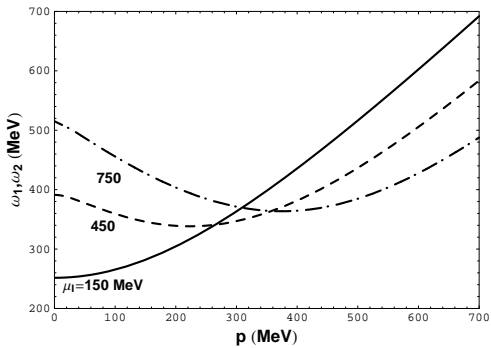}
\caption{The dispersion relation of the fermionic excitation
$\omega_1=\omega_2$ at $\mu_I=150,450,750$ MeV. \label{fig3}}
\end{center}
\end{figure}

To explain why there will be a BEC to BCS crossover, we focus
firstly on the dispersion relations of the fermionic excitations,
\begin{eqnarray}
&&\omega_{1,2}(p)=\sqrt{\left(\sqrt{p^2+m^2}-\mu_I/2\right)^2+\Delta^2},\nonumber\\
&&\omega_{3,4}(p)=\sqrt{\left(\sqrt{p^2+m^2}+\mu_I/2\right)^2+\Delta^2}.
\end{eqnarray}
In the case of $\mu_I>0$, $\omega_3=\omega_4$ is the anti-particle
excitation in the sense of isospin and irrelevant for our
discussion. In Fig.\ref{fig3} we showed the dispersion
$\omega_1=\omega_2$ at $\mu_I=150,\ 450,\ 750$ MeV. Obviously, the
fermionic excitations are always gapped. At small $\mu_I$ with
$\mu_I/2<m(\mu_I)$, the minimum of the dispersion is at $p=0$
where the energy gap is $\sqrt{\mu_N^2+\Delta^2}$ with
$\mu_N=\mu_I/2-m$ which can be regarded as the corresponding
non-relativistic chemical potential we will explain below.
However, at large $\mu_I$ with $\mu_I/2>m(\mu_I)$, the minimum of
the dispersion is shifted to $p\simeq\mu_I/2$ where the energy gap
is $\Delta$. This phenomenon is quite similar to the study of
nucleon dispersion at finite isospin density in the nonlinear
sigma model\cite{michae}. Such a phenomenon is a signal of BEC-BCS
crossover.
\begin{figure}[!htb]
\begin{center}
\includegraphics[width=6.5cm]{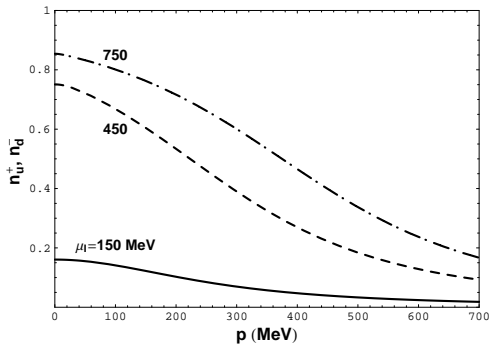}
\caption{The occupation number of the fermionic excitation
$n_u^+=n_d^-$ at $\mu_I=150,450,750$ MeV. \label{fig4}}
\end{center}
\end{figure}

Secondly, we discuss the occupation number of each kind of quarks
,
\begin{eqnarray}
&&n_u^+(p) = n_d^-(p) = v_-^2(p), \nonumber\\
&&n_d^+(p) = n_u^-(p) = v_+^2(p).
\end{eqnarray}
In Fig.\ref{fig4} we show the occupation number $n_u^+=n_d^-$ at
$\mu_I=150,\ 450,\ 750$ MeV. At small $\mu_I$ such as $\mu_I=150$
MeV the occupation number is very small and smooth in the whole
momentum region, while at large $\mu_I$ the occupation number
becomes large near $p=0$ and drops down with increasing momentum.
When $\mu_I$ becomes very large, the occupation number is indeed
of BCS type. This is the second signal of BEC-BCS crossover.

Finally, we should point out some similarity between such a
BEC-BCS crossover and the non-relativistic system. For small
$\mu_I$ where $m$ is large, the minimum of the dispersion is
located at $p=0$. Near this minimum with $p\ll m$, we can do the
non-relativistic expansion
\begin{equation}
\sqrt{p^2+m^2}\simeq m+\frac{p^2}{2m}+\cdots,
\end{equation}
and the corresponding non-relativistic chemical potential is
\begin{equation}
\mu_N=\frac{\mu_I}{2}-m.
\end{equation}
It is well known that for a non-relativistic Fermi gas, the
chemical potential will become negative in the BEC region.
Therefore, we can estimate the BEC region in our model by
requiring $\mu_{N}<0$,
\begin{equation}
\frac{\mu_I}{2}<m(\mu_I)=m(0)\left(\frac{m_\pi}{\mu_I}\right)^2,
\end{equation}
which determines another critical isospin chemical potential
$\mu_I^0$ shown in the phase diagram in the introduction,
\begin{equation}
\mu_I^0=\left[2m(0)m_\pi^2\right]^{1/3}.
\end{equation}
With the above chosen parameter set, we have $\mu_I^0=230$ MeV.
When the effective quark mass in vacuum varies from $300$ MeV to
$500$ MeV, $\mu_I^0$ changes in the region $230 - 270$ MeV. Such a
non-relativistic BEC in relativistic fermion superfluid was
investigated in detail in \cite{RBEC}. The critical value
$\mu_I^0$ we obtained here is similar to the one
$\mu_I^0\sim\left(m_nm_\pi^2\right)^{1/3}$ obtained in nonlinear
sigma model\cite{michae} where $m_n$ is the nucleon mass. Even
though the above discussion is only qualitative, it will be
important for us to understand the phase structure when a nonzero
baryon density is turned on.

The critical value $\mu_I^0$ has another physical meaning. The
chiral perturbation theory predicts that the chiral and pion
condensates can be related to each other by a chiral
rotation\cite{son}, i.e., for $\mu_I\geq m_\pi$ there are
\begin{equation}
\sigma(\mu_I)=\sigma(0)\cos\alpha,\ \ \
\pi(\mu_I)=\sigma(0)\sin\alpha
\end{equation}
with $\cos\alpha=m_\pi^2/\mu_I^2$. We found that this behavior of
chiral rotation holds approximately in the region
$m_\pi<\mu_I<\mu_I^0$ and breaks down for $\mu_I>\mu_I^0$ in the
NJL model.

\section {Sarma Phase at Finite $\mu_B$}
\label{s4}
When a baryon chemical potential is turned on, we may expect that
the pion condensate will jump to zero at a critical baryon
chemical potential which indicates a first order phase transition,
like the behavior of chiral condensate. Such an assumption has
been used in the studies of QCD phase diagram in the NJL
model\cite{bard,warringa}. However, when a possible Sarma phase is
taken into account, this will not be true. To see why a Sarma
phase can appear, let us compare the dispersion relations of the
fermionic excitations $\omega_i,\ i=1,2,3,4$ obtained here with
those obtained in the study of color superconductor with charge
neutrality\cite{CSC1}. The main differences between the two cases
are 1) the averaged Fermi surface of the two paired fermions is
controlled by baryon chemical potential in color superconductivity
but by isospin chemical potential in pion superfluidity, and the
mismatch is served by isospin chemical potential in color
superconductivity but by baryon chemical potential in pion
superfluidity, and 2) the effective quark mass is very small and
can be neglected in color superconductivity but plays an important
role in pion superfluidity when the isospin chemical potential is
not too large. Taking into account these two differences, we can
obtain from the conclusion for gapless color
superconductivity\cite{CSC2} that, the Sarma phase can be realized
in pion condensed matter when the gap $\Delta$ is smaller than one
third of the baryon chemical potential, $\Delta<\mu$.

From the possible solution of the gapless excitations, $\omega_2(
p)=0$ and $\omega_4(p)=0$, we derive the two zero points at
momenta $p_1$ and $p_2$ satisfying $p_1<p_2$,
\begin{equation}
p_1=m\sqrt{\lambda_1^2-1},\ \ \ p_2=m\sqrt{\lambda_2^2-1},
\end{equation}
with
\begin{eqnarray}
\label{standard}
\lambda_1&=&\frac{\mu_I/2-\sqrt{\mu^2-\Delta^2}}{m},\nonumber\\
\lambda_2&=&\frac{\mu_I/2+\sqrt{\mu^2-\Delta^2}}{m}.
\end{eqnarray}
Obviously, for  $\Delta>\mu$ or $\Delta<\mu$ but
$|\lambda_{1,2}|<1$, there is no real solution of the gapless
excitation, all branches of quasiparticles are gapped. In this
case there are
\begin{eqnarray}
&&n_u ^+(p) = n_d^-(p) = v_-^2, \nonumber\\
&&n_d^+(p) = n_u ^-(p) = v_+^2
\end{eqnarray}
for all $p$, and $n_B=0$. Only in the case with $\Delta<\mu$ there
is the possibility to realize the Sarma phase. According to the
values of $\lambda_{1,2}$, there are three possible types of Sarma
state.

{\bf Type 1}: $\lambda_1>0$ and $|\lambda_{1,2}|>1$. In this case,
only the branch $\omega_2$ has two gapless nodes $p_1$ and $p_2$,
the other branches are all gapped. This case is similar to the
gapless two flavor color superconductivity\cite{CSC2}. In this
phase we have
\begin{eqnarray}
\label{type1}
&&n_u ^+(p) = n_d^-(p) = v_-^2\ ,\nonumber\\
&&n_d^+(p) = n_u ^-(p) = v_+^2\
\end{eqnarray}
for $p<p_1$ and $p>p_2$ and
\begin{eqnarray}
\label{type2}
&&n_u ^+(p) =1, \ \ n_d^-(p) = 0\ , \nonumber\\
&&n_d^+(p) = n_u ^-(p) = v_+^2\
\end{eqnarray}
for $p_1<p<p_2$. The baryon number density is
\begin{equation}
n_B=\frac{p_2^3-p_1^3}{3\pi^2}=\frac{(\lambda_2^2-1)^{3/2}-(\lambda_1^2-1)^{3/2}}{3\pi^2}m^3,
\end{equation}
the nonzero baryon number comes from the normal component in the
region $p_1<p<p_2$.

{\bf Type 2}: $|\lambda_1|<1$ and $|\lambda_2|>1$. In this case,
only the branch $\omega_2$ has one gapless node $p_2$, the other
branches are all gapped. The system is in the paired state
(\ref{type1}) at $p>p_2$ and the state (\ref{type2}) without
pairing at $0<p<p_2$. The baryon number density becomes
\begin{equation}
n_B=\frac{p_2^3}{3\pi^2}=\frac{(\lambda_2^2-1)^{3/2}}{3\pi^2}m^3.
\end{equation}

{\bf Type 3}: $\lambda_1<0$ and $|\lambda_{1,2}|>1$. In this case,
the branch $\omega_2$ has a gapless node $p_2$ and the branch
$\omega_4$ has a gapless node $p_1$, the other branches are all
gapped. In this phase we have
\begin{equation}
n_u ^+(p) = n_d^-(p) = v_-^2\
\end{equation}
for $p>p_2$ and
\begin{equation}
n_u ^+(p) =1, \ \ n_d^-(p) = 0\
\end{equation}
for $0<p<p_2$, and
\begin{equation}
n_d^+(p) = n_u ^-(p) = v_+^2\
\end{equation}
for $p>p_1$ and
\begin{equation}
n_d^+(p) =1,\ \ n_u ^-(p) = 0\
\end{equation}
for $0<p<p_1$. The baryon number density is
\begin{equation}
n_B=\frac{p_1^3+p_2^3}{3\pi^2}=\frac{(\lambda_2^2-1)^{3/2}+(\lambda_1^2-1)^{3/2}}{3\pi^2}m^3.
\end{equation}

Before making numerical calculations, we firstly estimate where
the Sarma phase is of type $2$ only. From the judgement
(\ref{standard}), it is clear that the condition to have only type
2 is $m(\mu_I,\mu_B)>\mu_I/2$. Since the effective quark mass at
finite baryon density can not be larger than its value at zero
baryon density, we have
\begin{equation}
\mu_I<\mu_I^0,
\end{equation}
where $\mu_I^0$ is the critical isospin chemical potential we
obtained in last section.

Numerically solving the coupled gap equations (\ref{gap0}) for $m$
and $\Delta$, we can find all possible solutions, including the
gapped and gapless (Sarma) states. At a fixed isospin chemical
potential $\mu_I$, there is always a gapped solution
$\Delta(\mu_I,\mu_B)=\Delta(\mu_I,\mu_B=0)\equiv \Delta_0$
provided $\mu$ is less than the energy gap. Different from the
result obtained in \cite{ebert} where the gapless state can
survive only in a narrow isospin chemical potential region, we
found that the Sarma state can exist at any $\mu_I>m_\pi$.
Obviously, the gap $\Delta$ in Sarma state is smaller than the gap
$\Delta_0$ in gapped state due to the Fermi surface mismatch.
\begin{figure}[!htb]
\begin{center}
\includegraphics[width=6.5cm]{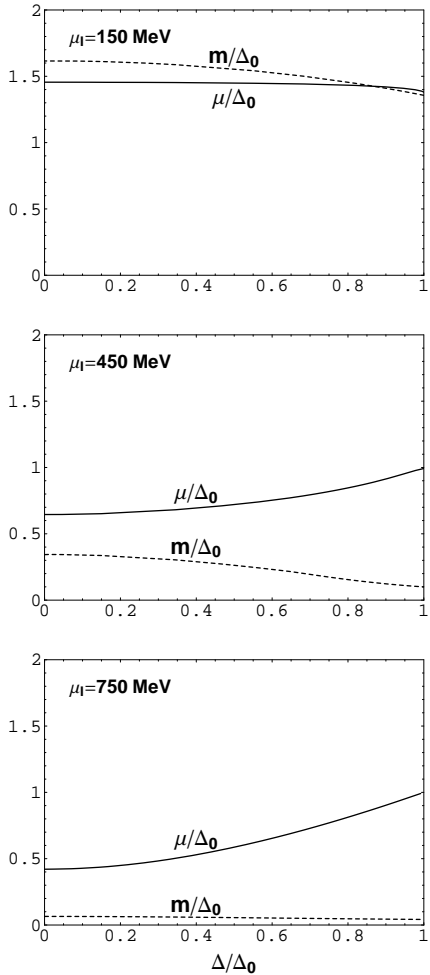}
\caption{The effective quark mass $m$ and baryon chemical
potential $\mu$ as functions of the scaled pion condensate
$\Delta/\Delta_0$ at $\mu_I=150,\ 450,\ 750$ MeV in the gapless
region. \label{fig5}}
\end{center}
\end{figure}
\begin{figure}[!htb]
\begin{center}
\includegraphics[width=6.5cm]{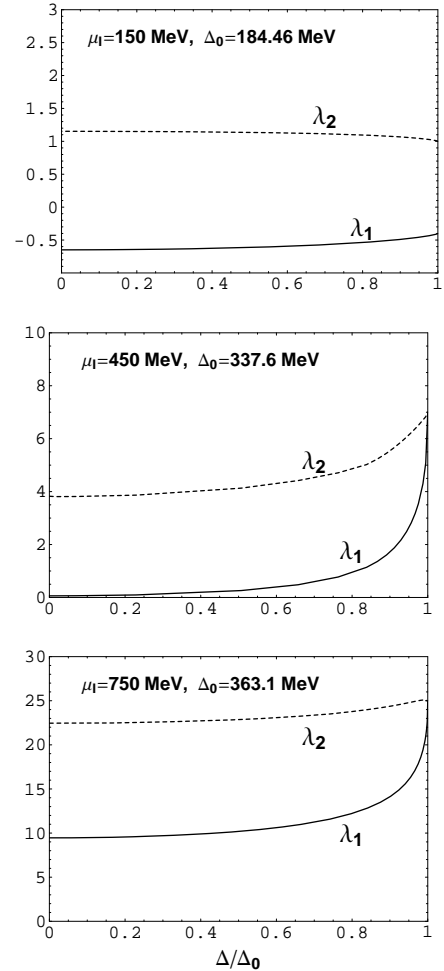}
\caption{The dimensionless quantities $\lambda_1$ and $\lambda_2$
determining the type of Sarma phase as functions of the scaled
pion condensate $\Delta/\Delta_0$ at $\mu_I=150,\ 450,\ 750$ MeV
in the gapless region. \label{fig6}}
\end{center}
\end{figure}

The effective quark mass $m$ and gap $\Delta$ in Sarma state are
shown in Fig.\ref{fig5} at different isospin chemical potentials.
To show only the Sarma state, we have restricted the condensate in
the region $0<\Delta/\Delta_0 <1$. For $\mu_I<\mu_I^0$ such as
$\mu_I=150$ MeV, the system is still in chiral breaking phase with
large effective quark mass, and the baryon chemical potential
$\mu$ starts at a value larger than $\Delta_0$ and then decreases
slowly with increasing $\Delta$ in the whole gapless region. The
behavior that $\mu$ decreases with $\Delta$ is important, since it
is directly related to the stability of the Sarma state, like the
case in the BEC region of asymmetric fermion
superfluids\cite{pao}. For $\mu_I>\mu_I^0$ such as $\mu_I=450$ and
$750$ MeV, the chiral symmetry is almost restored with very small
effective quark mass, and the baryon chemical potential $\mu$
starts at a value smaller than $\Delta_0$ and then goes up in the
gapless region.

The quantities $\lambda_1$ and $\lambda_2$ which determine the
type of the Sarma state are shown in Fig.\ref{fig6} as functions
of the scaled condensate $\Delta/\Delta_0$ at fixed isospin
chemical potential. The gapless solution is of type 2 at
$\mu_I<\mu_I^0$, as we analytically discussed above, and type 1 at
large enough $\mu_I$. For intermediate $\mu_I$, there exist both
type 1 at large $\Delta$ and type 2 at small $\Delta$. We never
found a solution of type 3. In fact, this type of Sarma phase
appears only in chiral limit where the effective quark mass is
always zero in pion condensation region\cite{ebert}.

\begin{figure}[!htb]
\begin{center}
\includegraphics[width=6.5cm]{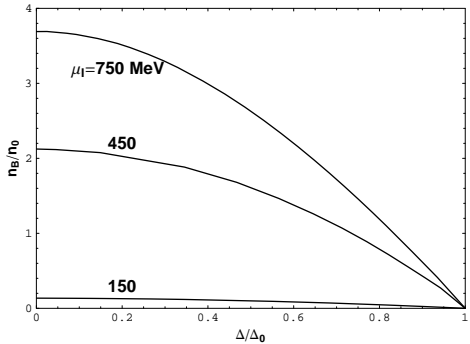}
\caption{The baryon number density $n_B$ in the gapless phase,
scaled by $n_0=0.16fm^{-3}$, as a function of $\Delta/\Delta_0$ at
$\mu_I=150,450,750MeV$. \label{fig7}}
\end{center}
\end{figure}

Because of the Fermi surface mismatch in the gapless state, the
baryon number density of the system is nonzero. In Fig.\ref{fig7}
we showed the baryon density $n_B$ as a function of
$\Delta/\Delta_0$. It is easy to understand that $n_B$ drops down
monotonously with increasing $\Delta$, namely with decreasing
degree of mismatch, and goes up with increasing isospin chemical
potential, namely with increasing average Fermi surface. Note that
the baryon density $n_B$ and the ratio of $n_B$ to the isospin
density $n_I$ here correspond, respectively, to the density
difference and the density asymmetry $\alpha$ in asymmetric
fermion superfluids\cite{pao,caldas,he-4}, and the ratio will
decrease with increasing isospin chemical potential.

\section {Thermodynamic Stability}
\label{s5}
\begin{figure}[!htb]
\begin{center}
\includegraphics[width=6.5cm]{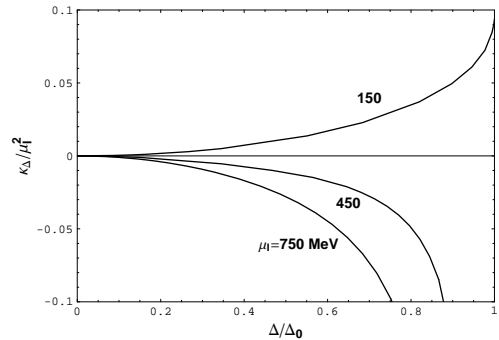}
\caption{The scaled susceptibility $\kappa_\Delta/\mu_I^2$ as a
function of the scaled pion condensate $\Delta/\Delta_0$ at
$\mu_I=150,\ 450,\ 750$ MeV. \label{fig8}}
\end{center}
\end{figure}

\begin{figure}[!htb]
\begin{center}
\includegraphics[width=6.5cm]{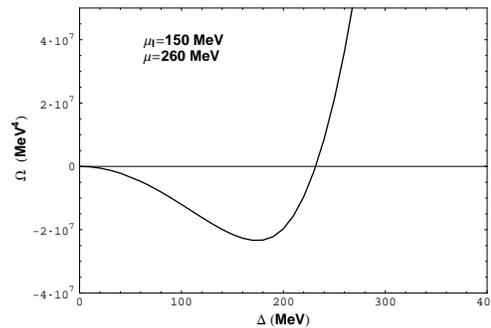}
\caption{The thermodynamic potential $\Omega$ as a function of the
pion condensate $\Delta$ at fixed chemical potentials $\mu=260$
MeV and $\mu_I=150$ MeV. \label{fig9}}
\end{center}
\end{figure}

To ensure that the Sarma phase we observed in the pion condensed
matter is stable, we should check both the thermodynamic and
dynamical stabilities of the system\cite{huang,pao}. In this
section we discuss the thermodynamic stability of the Sarma phase,
i.e., the stability against a small fluctuation of the order
parameter $\Delta$. To this end, we expand the thermodynamic
potential in powers of a small fluctuation $\delta$
\begin{equation}
\Omega(\Delta+\delta)-\Omega(\Delta)=\frac{\partial
\Omega}{\partial \Delta}\delta+\frac{1}{2}\frac{\partial^2
\Omega}{\partial \Delta^2}\delta^2+\cdots.
\end{equation}
The linear term vanishes automatically due to the gap equation
$\partial\Omega/\partial\Delta=0$. At zero temperature, the
quantity $\kappa_\Delta\equiv\frac{1}{6}\frac{\partial^2
\Omega}{\partial \Delta^2}$ which reflects the thermodynamic
stability can be directly evaluated as
\begin{equation}
\kappa_\Delta = \int{d^3{\bf p}\over (2\pi)^3}\left[{\Delta^2\over
(E_\Delta^-)^2}\left({\theta(\omega_2)\over E_\Delta^-}
-\delta(\omega_2)\right)+(\mu_I\rightarrow -\mu_I)\right],
\end{equation}
where $\theta(x)$ is the step function of $x$. In Fig.\ref{fig8}
we showed the scaled susceptibility $\kappa_\Delta/\mu_I^2$ in the
Sarma state as a function of the scaled pion condensate
$\Delta/\Delta_0$ at several values of $\mu_I$. At
$\mu_I<\mu_I^0$, $\kappa_\Delta$ is always positive in the whole
gapless region, which means that the Sarma state is free from the
Sarma instability. In Fig.\ref{fig9}, we plotted the thermodynamic
potential at fixed baryon chemical potential as a function of the
condensate, the minimum corresponds really to the Sarma state. As
is well known, to have a stable Sarma state at fixed chemical
potential, the system should be under some specific conditions,
for instance, a proper finite range interaction between the two
species of fermions with large mass difference. Here we found a
stable Sarma state at fixed chemical potential without mass
difference in the frame of a four-fermion point interaction. Such
a phenomenon was observed in a strong coupling model\cite{kita}
and is consistent with the argument in\cite{pao,son2}.
\begin{figure}[!htb]
\begin{center}
\includegraphics[width=6.5cm]{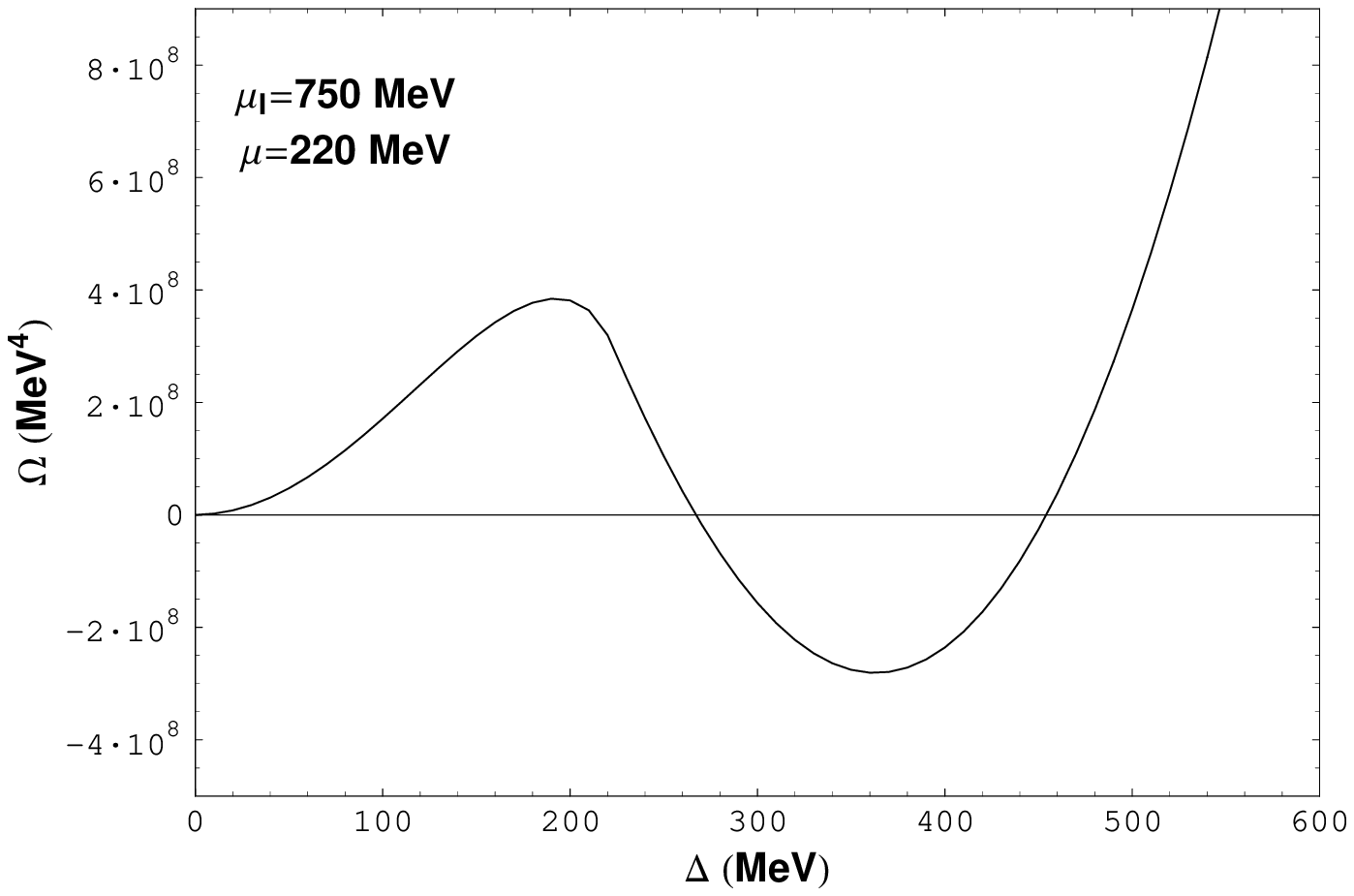}
\includegraphics[width=6.5cm]{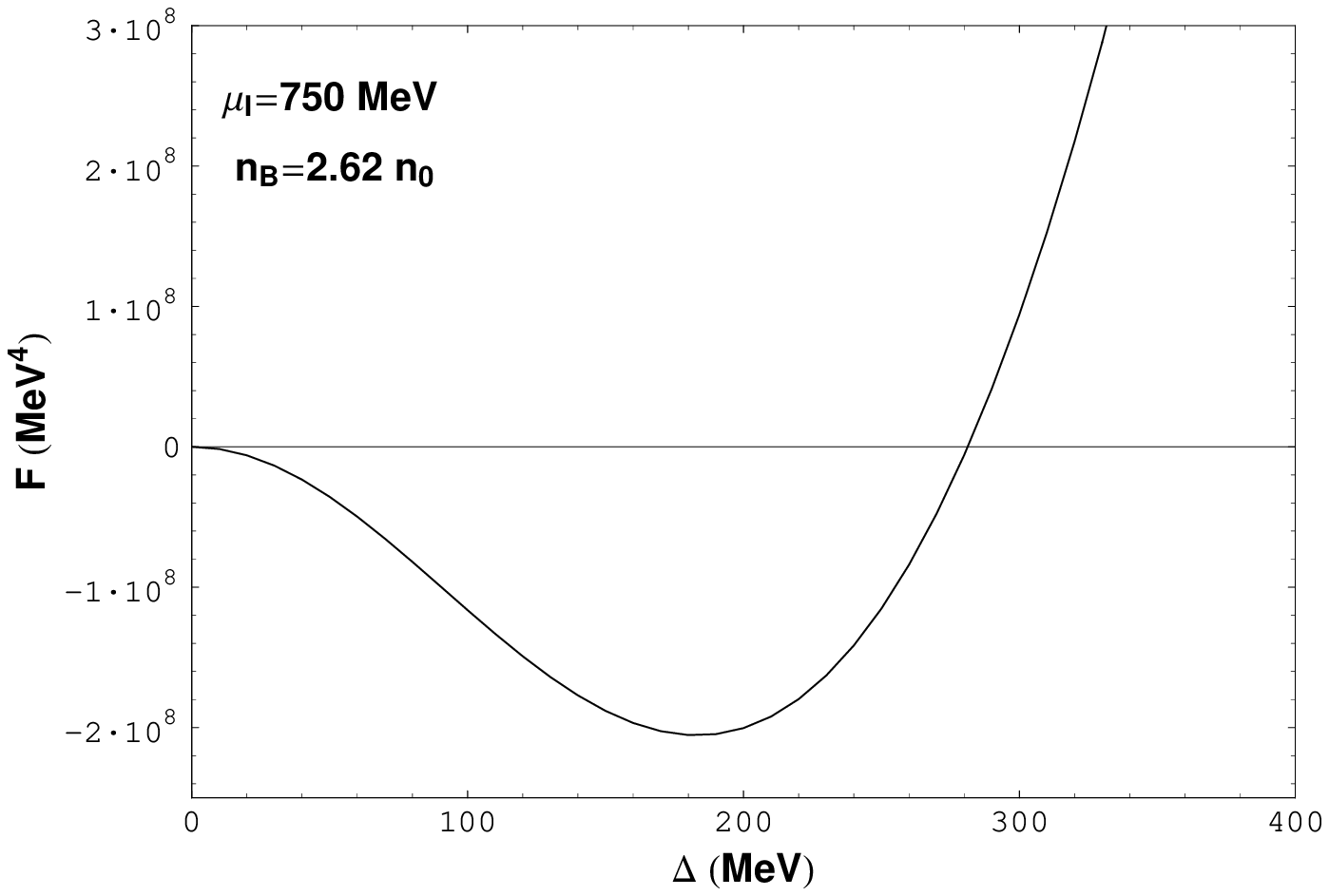}
\caption{The thermodynamic potential at fixed baryon chemical
potential and the free energy at fixed baryon number density as
functions of the pion condensate. \label{fig10}}
\end{center}
\end{figure}

For $\mu_I>\mu_I^0$, the susceptibility becomes negative and the
Sarma state suffers from the thermodynamic instability. The
question is, can we cure the Sarma instability at large $\mu_I$ by
considering baryon number conservation in physical systems? When
the baryon number is fixed to be nonzero which is required by
charge neutrality, the gapped phase is ruled out, and the possible
homogeneous and isotropic ground states are only normal phase
without pion condensation and Sarma phase. In this case, the Sarma
state may become stable like the gapless two flavor color
superconductivity\cite{CSC2}. Different from the system at fixed
baryon chemical potential where the ground state is determined by
the thermodynamic potential $\Omega$, the free energy
$F=\Omega+\mu_Bn_B$ controls the ground state for the system at
fixed baryon number density $n_B$. As a comparison, we plotted in
Fig.\ref{fig10} the thermodynamic potential at a fixed baryon
chemical potential $\mu=220$ MeV and the free energy $F$ at the
corresponding baryon number density $n_B=2.62n_0$ in the Sarma
state. While the Sarma state corresponds to the maximum of the
thermodynamic potential and is therefore unstable in the case at
fixed chemical potential, the minimum of the free energy is
located at the Sarma solution and the Sarma state is really stable
in the case at fixed baryon number density.

\section {Dynamical Stability}
\label{s6}
We discuss now the dynamical stability of the Sarma phase, i.e.,
the stability against the perturbation of an external current.
Since a non-zero expectation value
$\langle\bar{u}i\gamma_5d\rangle$ breaks both electromagnetic
$U_{em}(1)$ gauge symmetry and the global $U_I(1)$ symmetry, we
will discuss the responses of the system to the external
electromagnetic current
$J_{em}^\mu=e\bar{\psi}\hat{Q}\gamma^\mu\psi$ with
$\hat{Q}=diag(Q_u,Q_d)$ and to the isospin current
$J_3^\mu=\frac{1}{2}\bar{\psi}\tau_3\gamma^\mu\psi$. The former is
just the electromagnetic Meissner effect, and the later
corresponds to the superfluid density which is important to judge
the stability of the Sarma phase\cite{wu}.

\subsection {Response to Electromagnetic Current }
The quantity which describes the response of the system to an
external electromagnetic current is the photon self-energy or
photon polarization tensor defined by
\begin{equation}
\Pi^{\mu\nu}_{em}(p)=-\frac{i}{2}\int\frac{d^4k}{(2\pi)^4}\text
{Tr}\left[\hat{\Gamma}_{em}^\mu{\cal
S}(k)\hat{\Gamma}_{em}^\nu{\cal S}(k-p)\right]
\end{equation}
with $\hat{\Gamma}_{em}^\mu=e\hat{Q}\gamma^\mu$ at one loop level.
In the following we discuss only the long-wave and static property
of the response. The quantities that can describe this property
are the Debye mass $m_D$ and Meissner mass $m_M$ defined as
\begin{eqnarray}
&&m_D^2=-\lim_{{\bf p}\rightarrow0}\Pi_{em}^{00}(\omega=0,{\bf p}),\nonumber\\
&&m_M^2=-\frac{1}{2}\lim_{{\bf
p}\rightarrow0}(g_{ij}+\hat{p}_i\hat{p}_j)\Pi_{em}^{ij}(\omega=0,{\bf
p}).
\end{eqnarray}
Since there is no possible instability associated with the Debye
mass, we focus on the Meissner mass only. After some algebras, the
Meissner mass can be expressed in terms of the matrix elements of
the propagator ${\cal S}$,
\begin{equation}
m_M^2=\frac{3e^2}{2}\int\frac{d^3{\bf p}}{(2\pi)^3}\Big(Q_u^2{\cal
T}_{uu}+Q_d^2{\cal T}_{dd}+2Q_uQ_d{\cal T}_{ud}\Big)
\end{equation}
with the definition
\begin{eqnarray}
{\cal T}_{uu}&=&T\sum_n \text {Tr}\Big({\cal S}_{uu}\gamma_i{\cal
S}_{uu}\gamma_i\Big),\nonumber\\
{\cal T}_{dd}&=&T\sum_n \text {Tr}\Big({\cal S}_{dd}\gamma_i{\cal
S}_{dd}\gamma_i\Big),\nonumber\\
{\cal T}_{ud}&=&T\sum_n \text {Tr}\Big({\cal S}_{ud}\gamma_i{\cal
S}_{du}\gamma_i\Big),
\end{eqnarray}
where we should take the summation over the index $i=1,2,3$. To
avoid the UV divergence, we should subtract the vacuum
contribution, which ensures that $m_M^2$ is zero when there is no
pion condensation. For $m_M^2>0$, the system exhibits diamagnetic
Meissner effect, otherwise the response is paramagnetic which
means a magnetic instability\cite{CI1}.
\begin{figure}[!htb]
\begin{center}
\includegraphics[width=6.5cm]{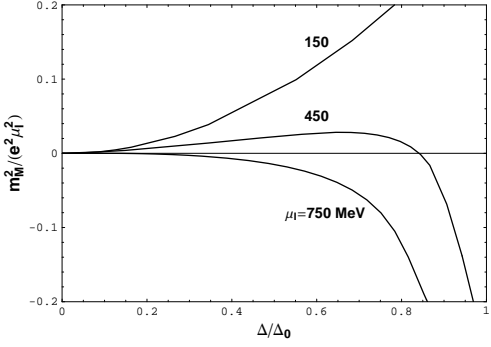}
\caption{The electromagnetic Meissner mass squared in the gapless
state, scaled by $e^2\mu_I^2$, as a function of the scaled pion
condensate $\Delta/\Delta_0$ at $\mu_I=150,\ 450,\ 750$ MeV.
\label{fig11}}
\end{center}
\end{figure}

Taking ${\cal T}_{uu}, {\cal T}_{dd}$ and ${\cal T}_{ud}$
calculated in Appendix \ref{app2}, we showed in Fig.\ref{fig11}
the Meissner mass squared $m_M^2$ for the gapless phase at
$\mu_I=150,\ 450,\ 750$ MeV. We found that $m_M^2$ is positive in
the whole gapless region for $\mu_I<\mu_I^0$ which means stable
Sarma state, and becomes negative in the whole region for large
enough $\mu_I$ such as $\mu_I=750$ MeV which indicates dynamical
instability of the Sarma state. At intermediate $\mu_I$ such as
$\mu_I=450$ MeV, the Sarma state is stable for small gap or large
baryon density and unstable for large gap or small baryon density.

\subsection {Response to Isospin Current }
The quantity which describes the response of the system to an
external isospin current is the isospin current-current
correlation tensor defined by
\begin{equation}
\Pi_{3}^{\mu\nu}(p)=-\frac{i}{2}\int\frac{d^4k}{(2\pi)^4}\text
{Tr}\left[\Gamma_{3}^\mu{\cal S}(k)\Gamma_{3}^\nu{\cal
S}(k-p)\right]
\end{equation}
with $\hat{\Gamma}_{3}^\mu=\frac{1}{2}\tau_3\gamma^\mu$. We still
focus on the long-wave and static property of the response by
considering the quantity $m_I^2$ defined as
\begin{equation}
m_I^2=-\frac{1}{2}\lim_{{\bf
p}\rightarrow0}(g_{ij}+\hat{p}_i\hat{p}_j)\Pi_{3}^{ij}(\omega=0,{\bf
p})
\end{equation}
which is proportional to the superfluid density $\rho_s$ of the
system. After some algebras, $m_I^2$ can be expressed in terms of
the matrix elements of the propagator ${\cal S}$,
\begin{equation}
m_I^2=\frac{3}{8}\int\frac{d^3{\bf p}}{(2\pi)^3}\Big({\cal
T}_{uu}+{\cal T}_{dd}-2{\cal T}_{ud}\Big).
\end{equation}
Similarly, we should also subtract the vacuum contribution to
ensure zero $m_I^2$ in the normal phase.

\begin{figure}[!htb]
\begin{center}
\includegraphics[width=6.5cm]{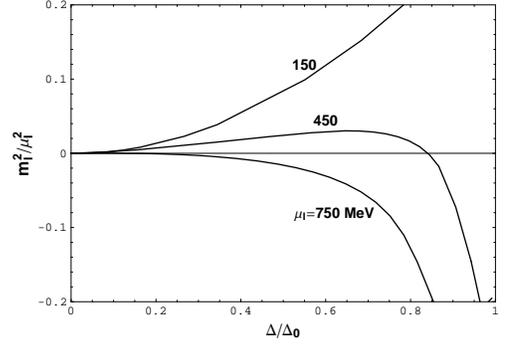}
\caption{The mass squared $m_I^2$ in the gapless phase, scaled by
$\mu_I^2$, as a function of the scaled pion condensate
$\Delta/\Delta_0$ at $\mu_I=150,\ 450,\ 750$ MeV. \label{fig12}}
\end{center}
\end{figure}

We calculated numerically the mass squared $m_I^2$ for the gapless
phase at $\mu_I=150,\ 450,\ 750$ MeV and showed the result in
Fig.\ref{fig12}. The $\mu_I$ dependence of the stability analysis
of the Sarma state is very similar to what shown in
Fig.\ref{fig11} for the response to the electromagnetic current.

In Figs.\ref{fig11} and \ref{fig12}, the negative $m_M^2$ and
$m_I^2$ behavior as $\mu(\mu^2-\Delta^2)^{-1/2}$ and become
divergent in the limit of $\Delta/\Delta_0\rightarrow 1$ where
$\mu\rightarrow\Delta$, as observed in gapless color
superconductivity\cite{CI1} and breached pairing
superfluidity\cite{he-3}. This divergence does not appear at small
$\mu_I<\mu_I^0$, since in the limit of $\mu\rightarrow\Delta$, the
gapless solution does not appear due to the fact that the zero
points of the excitations $p_{1,2}$ become imaginary when
$\mu_I/2<m$. This behavior is quite similar to the
model\cite{kita} proposed recently.

In conclusion, the Sarma pion condensed phase is always
thermodynamically stable at fixed baryon number density, and
dynamically stable at small isospin chemical potential
$\mu_I<\mu_I^0$ but dynamically unstable at large enough $\mu_I$.
For intermediate $\mu_I$ above and close to $\mu_I^0$, the Sarma
state is stable at large baryon density and unstable at small
baryon density.

\section {LOFF Pion Condensed Phase}
\label{s7}
The order parameters $\pi^+$ and $\pi^-$ are complex conjugate to
each other and can be set to be real only in a homogeneous and
isotropic superfluid. In this section we consider the possibility
that the phase factor $\theta$ in the order parameter is
non-uniform. Similar to the single plane wave LOFF state, we take
the following ansatz for the pion condensate
\begin{eqnarray}
\langle\bar{\psi}i\gamma_5\tau_+\psi\rangle &=&
\sqrt{2}\langle\bar{u}i\gamma_5d\rangle=\pi^+={\pi\over \sqrt
2}e^{2i{\bf q}\cdot{\bf x}},\nonumber\\
\langle\bar{\psi}i\gamma_5\tau_-\psi\rangle &=&
\sqrt{2}\langle\bar{d}i\gamma_5u\rangle=\pi^-={\pi\over \sqrt
2}e^{-2i{\bf q}\cdot{\bf x}}.
\end{eqnarray}
We call this phase with a nonzero wave vector ${\bf q}$ the LOFF
pion condensed phase. Note that this phase possesses the same
symmetry of the p-wave pion condensation\cite{migdal,sawyer,camp}
and is continued with the LOFF phase defined at ultra high
$\mu_I$\cite{son}. This phase is also similar to the chiral
crystal phase\cite{chiralLOFF}. After a transformation of the
quark fields for each flavors,
\begin{eqnarray}
\chi_u({\bf x})=u({\bf x})e^{-i{\bf q}\cdot{\bf x}},\ \ \
\chi_d({\bf x})=d({\bf x})e^{i{\bf q}\cdot{\bf x}},
\end{eqnarray}
the Lagrangian with the new fermion fields in mean field
approximation reads
\begin{eqnarray}
{\cal L} &=&
\bar{\chi}\left[i\gamma^\mu\partial_\mu-m+\hat{\mu}\gamma_0
-\tau_3{\bf \gamma}\cdot{\bf q}-i\Delta\tau_1\gamma_5\right]\chi \nonumber\\
&&-G(\sigma^2+\pi^2),
\end{eqnarray}
from which the inverse quark propagator matrix in the flavor space
as a function of quark momentum can be derived directly,
\begin{eqnarray}
&& {\cal S}^{-1}(p,{\bf q})=\\
&& \left(\begin{array}{cc} \gamma^\mu p_\mu-{\bf \gamma}\cdot{\bf q}+\mu_u\gamma_0-m & -i\gamma_5\Delta\\
-i\gamma_5\Delta & \gamma^\mu p_\mu+{\bf \gamma}\cdot{\bf
q}+\mu_d\gamma_0-m\end{array}\right),\nonumber
\end{eqnarray}
and the thermodynamic potential is then expressed as
\begin{equation}
\Omega=G(\sigma^2+\pi^2)-{T\over V}\ln \det {\cal S}^{-1}.
\end{equation}
To evaluate the thermodynamic potential, we employ the following
formula which can be derived from the formula listed in Appendix
\ref{app1},
\begin{widetext}
\begin{eqnarray}
\label{loff}
\left[\det{\cal S}^{-1}\right]^{1/2}
&=&\left[\left(p_0+\mu+\epsilon_-\right)^2-\left(\epsilon_+-\mu_I/2\right)^2-\Delta^2\right]
\left[\left(p_0+\mu-\epsilon_-\right)^2-\left(\epsilon_++\mu_I/2\right)^2-\Delta^2\right]\nonumber\\
&&+\ 2\Delta^2\left[{\bf p}^2+m^2-{\bf q}^2-\sqrt{|{\bf p}+{\bf
q}|^2+m^2}\sqrt{|{\bf p}-{\bf q}|^2+m^2}\right],
\end{eqnarray}
\end{widetext}
where $\epsilon_\pm$ are defined as
\begin{equation}
\epsilon_\pm=\frac{1}{2}\left(\sqrt{|{\bf p}+{\bf
q}|^2+m^2}\pm\sqrt{|{\bf p}-{\bf q}|^2+m^2}\right).
\end{equation}

The gap equations to determine simultaneously the condensates $m$
and $\Delta$ and the pair momentum ${\bf q}$ are derived from the
minimum of the thermodynamic potential, namely the equations
(\ref{gap}) and
\begin{equation}
\label{gap1}
{\partial\Omega\over\partial q}=0,\ \ \
{\partial^2\Omega\over\partial q^2}>0.
\end{equation}
Numerically solving the gap equations is not a easy task due to
the last term in (\ref{loff}). In the study of color
superconductivity, people simply dropped the last term since the
gap there is small but the baryon chemical potential is
large\cite{gorbar}. Here it seems not good to employ this
approximation unless $\mu_I$ is very large. In this paper we will
not directly calculate the LOFF state, but explain why the LOFF
pion condensed phase is favored at large $\mu_I$. Following the
treatment in \cite{he-4,ACI1}, we expand the thermodynamic
potential in powers of the pair momentum ${\bf q}$ in the vicinity
of ${\bf q}=0$,
\begin{equation}
\Omega(q)-\Omega(0)=\frac{\partial \Omega}{\partial
q}\Big|_{q=0}q+\frac{1}{2}\frac{\partial^2 \Omega}{\partial
q^2}\Big|_{q=0}q^2+\cdots.
\end{equation}
The linear term vanishes automatically because of the gap equation
(\ref{gap1}). Separating the pair momentum dependence of the
inhomogeneous quark propagator from the homogenous one,
\begin{equation}
{\cal S}^{-1}(p,{\bf q})={\cal S}^{-1}(p,{\bf q}=0)-\tau_3{\bf
\gamma}\cdot{\bf q},
\end{equation}
and taking the derivative expansion, we can easily obtain
\begin{equation}
\frac{\partial^2 \Omega}{\partial q^2}\Big|_{q=0}=m_I^2.
\end{equation}
Obviously, for $m_I^2>0$ at small $\mu_I$, the LOFF phase is less
favored than the Sarma phase, and for  $m_I^2<0$ at large $\mu_I$,
the LOFF phase is more favored than the Sarma phase.
\begin{figure}[!htb]
\begin{center}
\includegraphics[width=8cm]{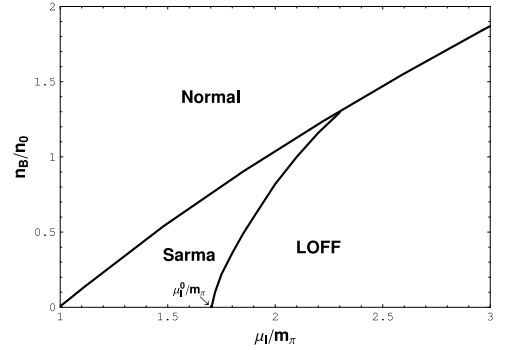}
\caption{The phase diagram of pion superfluidity in the
$\mu_I-n_B$ plane at zero temperature, calculated in the flavor
SU(2) NJL model. \label{fig13}}
\end{center}
\end{figure}

From the calculation on the pion superfluidity at fixed baryon
density in Section \ref{s4}, we can determine the phase transition
line from the pion superfluidity phase to the normal phase in the
$\mu_I-n_B$ plane at fixed temperature, and then with the
stability analysis discussed in Sections \ref{s5} and \ref{s6} and
in this section, the line which separates the inhomogeneous and
anisotropic LOFF state from the homogeneous and isotropic Sarma
state inside the pion superfluidity can be further fixed. The
phase diagram at zero temperature is shown in Fig.\ref{fig13}
which is very similar to the estimation shown in Fig.\ref{fig1}.

\section {Summary}
\label{s8}
As a first step to achieve a physical system with pion condensate,
we should consider the effect of a nonzero baryon density. In this
paper we have investigated the Sarma and LOFF pion condensed
matter at finite baryon number density in the frame of
Nambu-Jona-Lasinio model, and checked the thermodynamic and
dynamical stability of the Sarma phase.

The isospin chemical potential $\mu_I$ controls the averaged Fermi
surface of the quark-antiquark condensate, and the baryon chemical
potential $\mu_B$ offers a mismatch between the quark and
antiquark. At a fixed $\mu_I\ge\mu_I^c=m_\pi$, the magnitude of
the pion condensate decreases with increasing baryon density, and
finally vanishes at a critical baryon density where the system
undergoes a phase transition from the pion condensation to the
normal phase. The critical baryon density increases with
increasing isospin chemical potential. Inside the pion condensed
phase, the system goes from the Sarma state to the LOFF state when
$\mu_I$ increases. At low $\mu_I\le\mu_I^0$ where $\mu_I^0\sim
250$ MeV is determined by the pion mass and quark mass in the
vacuum, the coupling is strong and the effective quark mass is
large enough, the Sarma state is therefore thermodynamically and
dynamically stable. At high enough $\mu_I\gg\mu_I^0$, the
effective quark mass becomes very small, while the Sarma
instability can be avoided via fixing the baryon number density to
be nonzero, the dynamical instability of the Sarma state is still
there. In this case, the LOFF state is stable. In the intermediate
region with $\mu_I$ being above and close to $\mu_I^0$, the stable
ground state is LOFF state at lower baryon density and Sarma state
at higher baryon density. The above summary of the phase structure
of pion condensed matter is illustrated in Fig.\ref{fig13}.

Finally, we should point out that the diquark condensation in two
color QCD at finite baryon density has the same behavior as the
pion condensation in QCD with two or three colors at finite
isospin density. This has been confirmed by both lattice
simulation and low energy effective theory studies\cite{ratti}. We
expect that the same story will happen in two color QCD at finite
baryon and isospin chemical potentials.

{\bf Acknowledgement:} We thank Mei Huang, Judith A. McGovern,
Hai-cang Ren and Igor Shovkovy for helpful discussions and
comments. The work was supported in part by the grants
NSFC10428510, 10435080, 10575058 and SRFDP20040003103.

\begin{widetext}
\appendix
\section {Evaluating $\det {\cal S}^{-1}$}
\label{app1}
To evaluate the thermodynamic potential $\Omega$ at mean field
level, we should calculate the determinant of the inverse fermion
propagator with off-diagonal elements
\begin{equation}
{\cal S}^{-1}(p,{\bf q})=\left(\begin{array}{cc} {\cal
S}_1^{-1}(p,{\bf q})&-i\gamma_5\Delta
\\ -i\gamma_5\Delta&{\cal
S}_2^{-1}(p,{\bf q})\end{array}\right),
\end{equation}
where
\begin{equation}
{\cal S}_\alpha^{-1}(p,{\bf q})=(p_0+\mu_\alpha)\gamma_0-{\bf
\gamma}\cdot{\bf p}_\alpha-m_\alpha
\end{equation}
is the inverse propagator for the quasiparticles $\alpha=1,\ 2$
with the masses $m_1,\ m_2$, the chemical potentials $\mu_1,\
\mu_2$ and momenta ${\bf p}_1={\bf p}_2={\bf p}$ in isotropic
states and ${\bf p}_1={\bf p}+{\bf q},\ {\bf p}_2={\bf p}-{\bf q}$
in LOFF state. Taking into account of the identity
\begin{equation}
\det{\cal S}^{-1}=\det(\gamma^5{\cal S}^{-1}\gamma^5),
\end{equation}
we have
\begin{eqnarray}
\det{\cal S}^{-1}(p, {\bf q}) &=&
{\det}^{1/2}\left(\begin{array}{cc}
m_1^2-p_1^2-\Delta^2&(\gamma^\mu(p_1+p_2)_\mu-(m_1+m_2))i\gamma_5\Delta
\\
(\gamma^\mu(p_1+p_2)_\mu-(m_1+m_2))i\gamma_5\Delta&m_2^2-p_2^2-\Delta^2\end{array}\right)\nonumber\\
&=&
\left[(p_1^2-m_1^2+\Delta^2)(p_2^2-m_2^2+\Delta^2)-\Delta^2((p_1+p_2)^2-(m_1+m_2)^2)\right]^2
\end{eqnarray}
with $p_\alpha=(p_0+\mu_\alpha,{\bf p}_\alpha)$. In the simplest
case with $m_1=m_2=m$, ${\bf p}_1={\bf p}_2={\bf p}$ and
$\mu_1=-\mu_2=\mu$, we recover the BCS type result,
\begin{equation}
\det{\cal S}^{-1}(p)=\left(p_0^2-(\sqrt{{\bf
p}^2+m^2}-\mu)^2-\Delta^2\right)^2\left(p_0^2-(\sqrt{{\bf
p}^2+m^2}+\mu)^2-\Delta^2\right)^2.
\end{equation}

\section{Evaluating ${\cal T}_{uu}$, ${\cal T}_{dd}$ and ${\cal T}_{ud}$}
\label{app2}
We evaluate the quantities ${\cal T}_{uu}$, ${\cal T}_{dd}$ and
${\cal T}_{ud}$ in this Appendix. Firstly we calculate the trace
in the spin space. Using the relations
\begin{eqnarray}
&&\text {Tr}\left[\Lambda_\pm({\bf
p})\gamma^0\gamma^i\Lambda_\pm({\bf
p})\gamma^0\gamma^i\right]=2\frac{p^2}{E_p^2},\ \ \ \ \ \ \ \ \ \
\ \ \ \ \text
{Tr}\left[\Lambda_\pm({\bf p})\gamma^0\gamma^i\Lambda_\mp({\bf p})\gamma^0\gamma^i\right]=2\left(3-\frac{p^2}{E_p^2}\right),\nonumber\\
&&\text {Tr}\left[\Lambda_\pm({\bf
p})\gamma^5\gamma^i\Lambda_\pm({\bf
p})\gamma^5\gamma^i\right]=2\left(3-\frac{p^2}{E_p^2}\right),\ \ \
\text{Tr}\left[\Lambda_\pm({\bf
p})\gamma^5\gamma^i\Lambda_\mp({\bf
p})\gamma^5\gamma^i\right]=2\frac{p^2}{E_p^2}, \
\end{eqnarray}
we obtain
\begin{eqnarray}
\label{b2}
\frac{1}{N_c}\text{Tr}\Big({\cal S}_{uu}\gamma_i{\cal
S}_{uu}\gamma_i\Big)&=&2\frac{p^2}{E_p^2}\frac{(i\omega_n+\mu+E_p^-)^2}{[(i\omega_n+\mu)^2
-(E_\Delta^-)^2]^2}+2\frac{p^2}{E_p^2}\frac{(i\omega_n+\mu-E_p^+)^2}{[(i\omega_n+\mu)^2-(E_\Delta^+)^2]^2}\nonumber\\
&&+\
4\left(3-\frac{p^2}{E_p^2}\right)\frac{i\omega_n+\mu+E_p^-}{(i\omega_n+\mu)^2-
(E_\Delta^-)^2}\frac{i\omega_n+\mu-E_p^+}{(i\omega_n+\mu)^2-(E_\Delta^+)^2},\nonumber\\
\frac{1}{N_c}\text{Tr}\Big({\cal S}_{dd}\gamma_i{\cal
S}_{dd}\gamma_i\Big)&=&2\frac{p^2}{E_p^2}\frac{(i\omega_n+\mu-E_p^-)^2}{[(i\omega_n+\mu)^2-
(E_\Delta^-)^2]^2}+2\frac{p^2}{E_p^2}\frac{(i\omega_n+\mu+E_p^+)^2}{[(i\omega_n+\mu)^2-(E_\Delta^+)^2]^2}\nonumber\\
&&+\
4\left(3-\frac{p^2}{E_p^2}\right)\frac{i\omega_n+\mu-E_p^-}{(i\omega_n+\mu)^2-
(E_\Delta^-)^2}\frac{i\omega_n+\mu+E_p^+}{(i\omega_n+\mu)^2-(E_\Delta^+)^2},\nonumber\\
\frac{1}{N_c}\text{Tr}\Big({\cal S}_{ud}\gamma_i{\cal
S}_{du}\gamma_i\Big)&=&2\frac{p^2}{E_p^2}\frac{-\Delta^2}{[(i\omega_n+\mu)^2-(E_\Delta^-)^2]^2}+
2\frac{p^2}{E_p^2}\frac{-\Delta^2}{[(i\omega_n+\mu)^2-(E_\Delta^+)^2]^2}\nonumber\\
&&+\
4\left(3-\frac{p^2}{E_p^2}\right)\frac{-i\Delta}{(i\omega_n+\mu)^2-(E_\Delta^-)^2}\frac{-i\Delta}{(i\omega_n+\mu)^2-(E_\Delta^+)^2}.
\end{eqnarray}
Secondly we calculate the fermion frequency summations. From the
decomposition of the summation
\begin{equation}
T\sum_n\frac{(i\omega_n+\mu+\epsilon)^2}{[(i\omega_n+\mu)^2-E^2]^2}
=a+(E+\epsilon)^2 b+2\epsilon c
\end{equation}
for any constants $\epsilon$ and $E$, and the simple frequency
summations
\begin{eqnarray}
a &=& T\sum_n\frac{1}{(i\omega_n+\mu)^2-E^2}=\frac{f(E+\mu)+f(E-\mu)-1}{2E},\nonumber\\
b &=&
T\sum_n\frac{1}{[(i\omega_n+\mu)^2-E^2]^2}=\frac{1}{2E}\frac{\partial}{\partial
E}\left[\frac{f(E+\mu)+f(E-\mu)-1}{2E}\right],\nonumber\\
c &=& T\sum_n\frac{1}{(i\omega_n+\mu+E)^2(i\omega_n+\mu-E)} =
\frac{f(E+\mu)+f(E-\mu)-1}{4E^2}-\frac{1}{2E}\frac{\partial
f(E+\mu)}{\partial E}
\end{eqnarray}
with the definition $f^{\prime}(x)=\partial f(x)/\partial x$, we
can express some of the summations in (\ref{b2}) as
\begin{eqnarray}
T\sum_n\frac{(i\omega_n+\mu+E_p^-)^2}{[(i\omega_n+\mu)^2-(E_\Delta^-)^2]^2}
&=&u_-^2v_-^2\frac{f(\omega_1)+f(\omega_2)-1}{E_\Delta^-}+v_-^4f^\prime(\omega_1)+u_-^4f^\prime(\omega_2), \nonumber\\
T\sum_n\frac{(i\omega_n+\mu-E_p^-)^2}{[(i\omega_n+\mu)^2-(E_\Delta^-)^2]^2}
&=&u_-^2v_-^2\frac{f(\omega_1)+f(\omega_2)-1}{E_\Delta^-}+u_-^4f^\prime(\omega_1)+v_-^4f^\prime(\omega_2),
\nonumber\\
T\sum_n\frac{\Delta^2}{[(i\omega_n+\mu)^2-(E_\Delta^-)^2]^2}
&=&-u_-^2v_-^2\frac{f(\omega_1)+f(\omega_2)-1}{E_\Delta^-}+u_-^2v_-^2\left[f^\prime(\omega_1)+f^\prime(\omega_2)\right],\nonumber\\
T\sum_n\frac{(i\omega_n+\mu+E_p^+)^2}{[(i\omega_n+\mu)^2-(E_\Delta^+)^2]^2}
&=&u_+^2v_+^2\frac{f(\omega_3)+f(\omega_4)-1}{E_\Delta^+}+v_+^4f^\prime(\omega_3)+u_+^4f^\prime(\omega_4),
\nonumber\\
T\sum_n\frac{(i\omega_n+\mu-E_p^+)^2}{[(i\omega_n+\mu)^2-(E_\Delta^+)^2]^2}
&=&u_+^2v_+^2\frac{f(\omega_3)+f(\omega_4)-1}{E_\Delta^+}+u_+^4f^\prime(\omega_3)+v_+^4f^\prime(\omega_4), \nonumber\\
T\sum_n\frac{\Delta^2}{[(i\omega_n+\mu)^2-(E_\Delta^+)^2]^2}
&=&-u_+^2v_+^2\frac{f(\omega_3)+f(\omega_4)-1}{E_\Delta^+}+u_+^2v_+^2\left[f^\prime(\omega_3)+f^\prime(\omega_4)\right].
\end{eqnarray}
Consider the decomposition
\begin{equation}
T\sum_n\frac{i\omega_n+\mu+\epsilon_1}{(i\omega_n+\mu)^2-E_1^2}\frac{i\omega_n+\mu-\epsilon_2}{(i\omega_n+\mu)^2-E_2^2}
=d+(E_1-\epsilon_1)(E_2+\epsilon_2)e-(E_1-\epsilon_1)f-(E_2+\epsilon_2)g
\end{equation}
for any constants $\epsilon_1, \epsilon_2, E_1$ and $E_2$, and the
simple frequency summations
\begin{eqnarray}
d &=&
T\sum_n\frac{1}{i\omega_n+\mu-E_1}\frac{1}{i\omega_n+\mu-E_2}=\frac{1}{E_1-E_2}\left[f(E_1-\mu)-f(E_2-\mu)\right],\nonumber\\
e &=&
T\sum_n\frac{1}{(i\omega_n+\mu)^2-E_1^2}\frac{1}{(i\omega_n+\mu)^2-E_2^2}
=\frac{1}{E_1^2-E_2^2}\left[\frac{f(E_1+\mu)+f(E_1-\mu)-1}{2E_1}-\frac{f(E_2+\mu)+f(E_2-\mu)-1}{2E_2}\right],\nonumber\\
f &=&
T\sum_n\frac{1}{(i\omega_n+\mu)^2-E_1^2}\frac{1}{i\omega_n+\mu-E_2}
=\frac{f(E_2-\mu)}{E_2^2-E_1^2}+\frac{f(E_1-\mu)}{2E_1(E_1-E_2)}+\frac{1-f(E_1+\mu)}{2E_1(E_1+E_2)},\nonumber\\
g &=&
T\sum_n\frac{1}{i\omega_n+\mu-E_1}\frac{1}{(i\omega_n+\mu)^2-E_2^2}
=\frac{f(E_1-\mu)}{E_1^2-E_2^2}+\frac{f(E_2-\mu)}{2E_2(E_2-E_1)}+\frac{1-f(E_2+\mu)}{2E_2(E_2+E_1)},
\end{eqnarray}
we can express the other summations in (\ref{b2}) as
\begin{eqnarray}
&&T\sum_n\frac{i\omega_n+\mu+E_p^-}{(i\omega_n+\mu)^2-(E_\Delta^-)^2}\frac{i\omega_n+\mu-E_p^+}{(i\omega_n+\mu)^2-(E_\Delta^+)^2}\nonumber\\
&=&\frac{f(\omega_2)-f(\omega_4)}{E_\Delta^--E_\Delta^+}+\frac{(E_\Delta^--E_p^-)(E_\Delta^++E_p^+)}{(E_\Delta^-)^2-(E_\Delta^+)^2}\left[\frac{f(\omega_1)+f(\omega_2)-1}{2E_\Delta^-}-\frac{f(\omega_3)+f(\omega_4)-1}{2E_\Delta^+}\right]\nonumber\\
&&-(E_\Delta^--E_p^-)\left[\frac{f(\omega_4)}{(E_\Delta^+)^2-(E_\Delta^-)^2}+\frac{f(\omega_2)}{2E_\Delta^-(E_\Delta^--E_\Delta^+)}+\frac{1-f(\omega_1)}{2E_\Delta^-(E_\Delta^-+E_\Delta^+)}\right]\nonumber\\
&&-(E_\Delta^++E_p^+)\left[\frac{f(\omega_2)}{(E_\Delta^-)^2-(E_\Delta^+)^2}+\frac{f(\omega_4)}{2E_\Delta^+(E_\Delta^+-E_\Delta^-)}+\frac{1-f(\omega_3)}{2E_\Delta^+(E_\Delta^++E_\Delta^-)}\right]\ , \nonumber\\
&&T\sum_n\frac{i\omega_n+\mu-E_p^-}{(i\omega_n+\mu)^2-(E_\Delta^-)^2}\frac{i\omega_n+\mu+E_p^+}{(i\omega_n+\mu)^2-(E_\Delta^+)^2}\nonumber\\
&=&\frac{f(\omega_2)-f(\omega_4)}{E_\Delta^--E_\Delta^+}+\frac{(E_\Delta^-+E_p^-)(E_\Delta^+-E_p^+)}{(E_\Delta^-)^2-(E_\Delta^+)^2}\left[\frac{f(\omega_1)+f(\omega_2)-1}{2E_\Delta^-}-\frac{f(\omega_3)+f(\omega_4)-1}{2E_\Delta^+}\right]\nonumber\\
&&-(E_\Delta^-+E_p^-)\left[\frac{f(\omega_4)}{(E_\Delta^+)^2-(E_\Delta^-)^2}+\frac{f(\omega_2)}{2E_\Delta^-(E_\Delta^--E_\Delta^+)}+\frac{1-f(\omega_1)}{2E_\Delta^-(E_\Delta^-+E_\Delta^+)}\right]\nonumber\\
&&-(E_\Delta^+-E_p^+)\left[\frac{f(\omega_2)}{(E_\Delta^-)^2-(E_\Delta^+)^2}+\frac{f(\omega_4)}{2E_\Delta^+(E_\Delta^+-E_\Delta^-)}+\frac{1-f(\omega_3)}{2E_\Delta^+(E_\Delta^++E_\Delta^-)}\right]\ , \nonumber\\
&&T\sum_n\frac{\Delta}{(i\omega_n+\mu)^2-(E_\Delta^-)^2}\frac{\Delta}{(i\omega_n+\mu)^2-(E_\Delta^+)^2}\nonumber\\
&=&\frac{\Delta^2}{(E_\Delta^-)^2-(E_\Delta^+)^2}\left[\frac{f(\omega_1)+f(\omega_2)-1}{2E_\Delta^-}-\frac{f(\omega_3)+f(\omega_4)-1}{2E_\Delta^+}\right]\
.
\end{eqnarray}

For a general combination
\begin{equation}
{\cal
T}(\alpha,\beta)=\alpha^2{\cal T}_{uu}+\beta^2{\cal
T}_{dd}-2\alpha\beta{\cal T}_{ud},
\end{equation}
we can separate it into a diamagnetic part ${\cal T}_d$ and a
paramagnetic part ${\cal T}_p$,
\begin{equation}
{\cal T}(\alpha,\beta)=2\frac{p^2}{E_p^2}{\cal
T}_p+4(3-\frac{p^2}{E_p^2}){\cal T}_d
\end{equation}
with
\begin{eqnarray}
{\cal T}_p
&=&(\alpha-\beta)^2u_-^2v_-^2\frac{f(\omega_1)+f(\omega_2)-1}{E_\Delta^-}+
(\alpha v_-^2+\beta u_-^2)^2f^\prime(\omega_1)+(\alpha u_-^2+\beta v_-^2)^2f^\prime(\omega_2)\\
&&+(\alpha-\beta)^2u_+^2v_+^2\frac{f(\omega_3)+f(\omega_4)-1}{E_\Delta^+}+(\alpha
u_+^2+\beta v_+^2)^2f^\prime(\omega_3)+(\alpha v_+^2+\beta
u_+^2)^2f^\prime(\omega_4),\nonumber\\
{\cal
T}_d&=&\frac{1}{(E_\Delta^-)^2-(E_\Delta^+)^2}\left[\frac{2\alpha\beta\Delta^2+(\alpha^2+\beta^2)
((E_\Delta^-)^2-E_p^-E_p^+)}{2E_\Delta^-}(f(\omega_1)+f(\omega_2)-1)-(\mu_I\rightarrow -\mu_I)\right]\nonumber\\
&&+(\alpha^2-\beta^2)\frac{E_\Delta^-E_p^+-E_\Delta^+E_p^-}{(E_\Delta^-)^2-(E_\Delta^+)^2}\left[\frac{f(\omega_1)+f(\omega_2)-1}{2E_\Delta^-}-(\mu_I\rightarrow
-\mu_I)\right]\nonumber\\
&&+\frac{\alpha^2-\beta^2}{(E_\Delta^+)^2-(E_\Delta^-)^2}\left[E_p^-f(\omega_4)+\frac{E_p^-}{2}
\left(1+\frac{E_\Delta^+}{E_\Delta^-}\right)f(\omega_2)+\frac{E_p^-}{2}\left(1-\frac{E_\Delta^+}{E_\Delta^-}\right)(1-f(\omega_1))+(\mu_I\rightarrow
-\mu_I)\right].\nonumber
\end{eqnarray}
We set $\alpha=Q_u=2/3$ and $\beta=-Q_d=1/3$ for calculating
$m_M^2$ and $\alpha=\beta=1$ for $m_I^2$. Note that only for
$\alpha=\beta$, the formula is invariant under the exchange
$\mu_I\rightarrow -\mu_I$, otherwise the isospin symmetry is
explicitly broken.
\end{widetext}


\newpage
\begin{thebibliography}{99}
\small
 \bibitem{son}       D.T.Son and M.A.Stephanov, Phys.Rev.Lett.{\bf 86},592(2001);Phys.Atom.Nucl.{\bf 64},834(2001).
 \bibitem{migdal}    A.B.Migdal,Zh.Eksp.Teor.Fiz.61,2210(1971)(Sov.Phys.JETP 36, 1052(1973)).
 \bibitem{sawyer}    R.F.Sawyer, Phys.Rev.Lett.{\bf 29},382(1972); D.J.Scalapino,Phys.Rev.Lett.{\bf 29},386(1972).
 \bibitem{camp}      D.K.Campbell,R.F.Dashen and J.T.Manassah, Phys.Rev.{\bf D12},979(1975);{\bf D12},1010(1975).
 \bibitem{kaplan}    D.B.Kaplan and A.E.Nelson, Phys.Lett.{\bf B175},57(1986).
 \bibitem{kogut}     J.B.Kogut,D.K.Sinclair, Phys.Rev.{\bf D66}, 034505(2002); {\bf D66}, 014508(2002); {\bf D70}, 094501(2004).
 \bibitem{toub}      D.Toublan and J.B.Kogut, Phys.Lett. {\bf B564}, 212(2003).
 \bibitem{frank}     M.Frank, M.Buballa and M.Oertel, Phys.Lett. {\bf B562}, 221(2003).
 \bibitem{bard}      A.Barducci, R.Casalbuoni, G.Pettini, L.Ravagli, Phys.Rev. {\bf D69}, 096004(2004); {\bf D71}, 016011(2005).
 \bibitem{he-1}      L.He and P.Zhuang, Phys.Lett.{\bf B615}, 93(2005); L.He, M.Jin and P.Zhuang, hep-ph/0503249.
 \bibitem{he-2}      L.He, M.Jin and P.Zhuang, Phys.Rev.{\bf D71}, 116001(2005).
 \bibitem{warringa}  H.J.Warringa, D.Boer and J.O.Andersen, Phys.Rev.{\bf D72}, 014015(2005).
 \bibitem{ebert}     D.Ebert and K.G.Klimenko, J.Phys.{\bf G32}, 599(2006); hep-ph/0510222.
 \bibitem{sarma}     G.Sarma, J.Phys.Chem.Solid {\bf 24},1029(1963).
 \bibitem{liu}       W.V.Liu and F.Wilczek, Phys. Rev. Lett.{\bf 90},047002(2003).
 \bibitem{pao}       C.H.Pao, S.Wu and S.K.Yip, cond-mat/0506437.
 \bibitem{son2}       D.T.Son and M.A.Stephanov, cond-mat/0507586.
 \bibitem{np}        A.Sedrakian and U.Lombardo, Phys. Rev. Lett. {\bf 84}, 602(2000).
\bibitem{CSC1}       M.Huang, P.Zhuang, and W.Chao, Phys. Rev. {\bf D67},065015(2003).
\bibitem{CSC2}       I.Shovkovy and M.Huang, Phys. Lett. {\bf B564},205(2003).
\bibitem{CSC3}       M.Huang and I.Shovkovy, Nucl. Phys. {\bf A729},835(2003).
\bibitem{CSC4}       M.Alford, C.Kouvaris and K. Rajagopal, Phys. Rev. Lett. {\bf 92},222001(2004).
\bibitem{forbes}     M.M.Forbes, E.Gubankova, W. Vincent Liu, F.Wilczek, Phys. Rev. Lett.{\bf 94},017001(2005).
\bibitem{wu}         S.Wu and S.Yip, Phys. Rev. {\bf A67}, 053603(2003).
\bibitem{CI1}        M.Huang and I.Shovkovy, Phys.Rev. {\bf D70}, R051501(2004); {\bf D70}, 094030(2004).
\bibitem{he-3}       L.He, M.Jin and P.Zhuang, Phys.Rev. {\bf B73} 024511(2006).
\bibitem{LOFF1}      A.I.Larkin and Yu.N.Ovchinnikov, Sov.Phys. JETP {\bf 20},762(1965).
\bibitem{LOFF2}      P.Fulde and R.A.Ferrell,Phys. Rev {\bf 135},A550(1964).
\bibitem{ACI1}       I.Giannakis and H.Ren, Phys. Lett. {\bf B611}, 137(2005); Nucl. Phys. {\bf B723}, 255(2005).
\bibitem{ACI3}       I.Giannakis, D.Hou and H.Ren, Phys.Lett. {\bf B631} 16(2005).
\bibitem{he-4}       L.He, M.Jin and P.Zhuang,  Phys. Rev. {\bf B73}, 214527(2006).
\bibitem{LOFF3}      M.Alford, J.Bowers and K.Rajagopal, Phys.Rev. {\bf D63},074016(2001).
\bibitem{LOFF4}      R.Casalbuoni and G.Nardulli, Rev.Mod.Phys. {\bf 76}, 263(2004).
\bibitem{LOFF5}      A.Sedrakian, Phys.Rev. {\bf C63}, 025801(2001).
\bibitem{LOFF6}      H.M¨¹ther and A.Sedrakian, Phys.Rev. {\bf C67}, 015802(2003).
\bibitem{LOFF7}      A.Sedrakian, J.Mur-Petit, A.Polls and H.M¨¹ther, Phys.Rev. {\bf A72}, 013613(2005).
\bibitem{smit}       E.Gubankova, A.Schmitt and F.Wilczek, cond-mat/0603603.
\bibitem{kita}       M.Kitazawa, D.Rischke and A.Shovkovy, hep-ph/0602065.
\bibitem{michae}     Michael C. Birse, Thomas D. Cohen and Judith A. McGovern, Phys.Lett. {\bf B516}, 27(2001).
\bibitem{RBEC}       Y.Nishida and H.Abuki, Phys.Rev. {\bf D72}, 096004(2005).
\bibitem{caldas}     H.Caldas, Phys. Rev. {\bf A69}, 063602(2004).
\bibitem{huang}      M.Huang, Phys.Rev. {\bf D73}, 045007(2006).
\bibitem{chiralLOFF} M.Sadzikowski, Phys.Lett.{\bf B553},
45(2003).
\bibitem{gorbar}     E.V.Gorbar, M.Hashimoto and V.A.Miransky, Phys.Rev.Lett. {\bf 96}, 022005 (2006).
\bibitem{ratti}      C.Ratti and W.Weise, Phys.Rev.{\bf D70}, 054013(2004).
\end{thebibliography}
\end{document}